# THE ROLE OF INSTITUTIONS IN THE DESIGN OF COMMUNICATION TECHNOLOGIES

Rajiv C. Shah* & Jay P. Kesan**

ABSTRACT

Communication technologies contain embedded values that affect our society's fundamental values, such as privacy, freedom of speech, and the protection of intellectual property. Researchers have shown the design of technologies is not autonomous but shaped by conflicting social groups. Consequently, communication technologies contain different values when designed by different social groups. Continuing in this vein, we argue that the institutions where communication technologies are designed and developed are an important source of the values of communication technologies. We use the term code to collectively refer to the hardware and software of communication technologies.

Institutions differ in their motivations, structure, and susceptibility to external influences. First, we focus on the political, economic, social, and legal influences during the development of communication technologies. The institutional reactions to these influences are embodied in code. Second, we focus on the decision-making issues in the review process for code. This process determines the code's content and affects the dissemination of code through the decision whether to publicly release the code.

Our analysis focuses on four institutional environments: universities, firms, open source movement, and consortia, which have contributed to the development of the Internet. To study these institutions we chose four historical case studies. They are: NCSA Mosaic developed at the University of Illinois; cookies developed by Netscape; the Apache web server developed by the open source movement; and the Platform for Internet Content Selection (PICS) developed by the World Wide Web Consortium (W3C).

The varying influences on institutions results in different qualities for code. Some of our significant findings were that within a firm the primary influence is the anticipation of consumer demand. A consequence is that firms do not develop code that is deemed as unprofitable. For example, Netscape incorporated the cookies technology to support commerce. Its motivation was generating market share and not minimizing the potential privacy loss. The goals and structure of a consortium are heavily influenced by the demands of its members. For example, PICS was developed because the W3C's members sought to avoid regulation. Consequently, the

___________________________

* Doctoral Candidate, Institute of Communications Research, University of Illinois at Urbana-Champaign.
** Assistant Professor, College of Law and Institute of Government and Public Affairs, University of Illinois at Urbana-Champaign.

This material is based upon work supported by the National Science Foundation under Grant No. ITR-0081426. Any opinions, findings, and conclusions or recommendations expressed in this material are those of the authors and do not necessarily reflect the views of the National Science Foundation.

design of PICS helped to avoid regulation, but it does not provide a sustainable economic model to encourage the use of PICS.

The review process for code differed among institutions. The factors included the type of decision makers, the criteria for the decision, as well as the decision making process. For example, a university provides its researchers autonomy in deciding the code's content and when to release code. But a university expects the code to be publicly released. The open source movement utilizes a process that is transparent and allows anyone to publicly comment on its code. As a result, problems in the code can be quickly found and corrected. Thus, the institution where a technology is designed plays a crucial role in the values incorporated into a communications technology.

I. INTRODUCTION

The most compelling method of online regulation is not law but code.[1] Code consists of the hardware and software of communication technologies and is the architecture of cyberspace. It is code that defines and creates online worlds, such as the Internet, for communication, commerce, and education. Code is not neutral but affected by biases that favor certain groups or values.[2] For example, the code of search engines has been shown to favor wealthy and powerful web sites.[3]

To better understand how code regulates, we are conducting a number of historical case studies on code. Our first goal is to understand the origins of code. Our initial analysis found that different institutions design and develop technologies differently.[4] This paper analyzes the role of institutions in the design and development of code in two areas. First, we consider the influence of social, political, economic, and legal factors on the development of code. Institutions respond differently to these influences and the results are embodied in code. Secondly, we examine the different institutional decision-making processes for the review of code. Institutions differ on the criteria for disseminating code publicly as well as the role of the public in the process.

The institutional variations on the development of code and the review process leads to code carrying different social values depending upon the origins of its development. This paper

---

[1] LAWRENCE LESSIG, CODE AND OTHER LAWS OF CYBERSPACE (1999).
[2] Batya Friedman & Helen Nissenbaum, *Bias in Computer Systems*, 14 ACM TRANS. INFO. SYSTEMS 1330 (1996), *see also* BATYA FRIEDMAN, HUMAN VALUES AND THE DESIGN OF COMPUTER TECHNOLOGY (1997) (arguing the design of technologies favors or biases certain uses over others).
[3] Lucas Introna & Helen Nissenbaum, *Defining the Web: The Politics of Search Engines*, IEEE COMPUTER, Jan. 2000, at 54.
[4] Our analysis is focused on institutions and not individuals because in the aggregate it is the institutions that design cyberspace. Although designers are individuals, they work within institutions. They are subject to the rules and norms of these institutions, thus attenuating any individual desires. For example, a corporation by definition seeks to build products that lead to financial gain. Jane E. Fountain, *Constructing the Information Society: Women,*

seeks to understand this process and the inherent tendencies of certain institutions. These tendencies affect qualities of code such as the type of intellectual property protection, ease of use, and support for open standards. Our eventual goal is to develop normative proposals to ensure the development of innovative code to meet the needs of society.

This paper focuses on four institutions that have been important in the development of the Internet. First, we study the development of code within universities. Universities are an important source of innovation. The second institution is the firm. Firms are the primary developers of code and are driven by profits. The third institution is the open source movement, which is unique to the development of code. The open source movement consists of volunteer programmers that are devoted to keeping code accessible and free for the public. Finally, we consider consortia which are cooperative efforts usually consisting of firms.

To investigate these institutions we have conducted four case studies. The case studies include the development of the first popular web browser, NCSA Mosaic, and the most widely used web server, Apache. The other two case studies focus on cookies and the Platform for Internet Content Selection (PICS). Cookies allow web sites to maintain information about their visitors. While PICS is a standard for labeling web pages for the purpose of limiting access to inappropriate material.

This paper is organized as follows. Part II provides a short background on the four case studies. Part III discusses the social, economic, political, and legal influences on each institution. Part IV focuses on the review process for code. This is the decision-making criteria and processes an institutions follows to decide when and what code to release.

---

*Information Technology, and Design*, 22 TECH. & SOC'Y 45 (2001) (arguing the appropriate level of analysis is the institution in the development of code).

II. OVERVIEW OF CASE STUDIES

This part provides a short background into the four case studies. A forthcoming paper will provide more complete detail on each of these case studies. The first case study is NCSA Mosaic developed by a university. The second case study is cookies developed by Netscape a firm. The third case study is the web server Apache developed by the open source movement. Finally, there is the Platform for Internet Content Selection developed by a consortium.

*A. NCSA Mosaic*

The origins of the World Wide Web (web) occurred at the Conseil Europeen pour la Recherche Nucleaire (CERN). This is a laboratory for particle physics funded by twenty European countries. Tim Berners-Lee conceived the web in 1989 at CERN.[5] He envisioned the web as connecting disparate information sources. For example, the web at CERN would allow access to the telephone book, conference information, a remote library system, and help files.[6] By 1991, Berners-Lee and Robert Cailliau developed a browser and server software for the web.[7]

The next major step in growth of the web occurred at the National Center for Supercomputing Applications (NCSA) at the University of Illinois at Urbana-Champaign. In the fall of 1992, Marc Andreessen worked for Ping Fu on some visualization projects at NCSA. Ping Fu asked Andreessen to write a graphical interface for a browser. He replied, "What's a browser?" She then showed Andreessen an early hypermedia system with links. She asked Andreessen to develop a tool that would allow people to download software by just clicking on a button.  Andreessen, replied, "Isn't that hard code FTP?"  She answered, "Marc, you can do something more intelligent than that!"[8]

---

[5] JAMES GILLIES & ROBERT CAILLIAU, HOW THE WEB WAS BORN 183 (2000).
[6] TIM BERNERS-LEE, WEAVING THE WEB: THE ORIGINAL DESIGN AND ULTIMATE DESTINY OF THE WORLD WIDE WEB BY ITS INVENTOR 20 (1999).
GILLIES & CAILLIAU, *supra* note at 202-203.

Later in November 1992, Andreessen watched a demonstration of the web by NCSA staff member Dave Thompson.  Thompson thought the web might be an innovative solution for an online collaboration project.[9]  Andreessen was inspired by this demonstration and begin investigating the web through the www-talk newsgroup hosted by CERN.  Eventually, he began work on NCSA Mosaic after learning the shortcomings of the other early web browsers.  In less than a year, NCSA Mosaic was available for the Unix, Windows, and McIntosh operating systems.[10]  NCSA Mosaic rapidly become the de facto method of accessing the web and introduced millions of people to the web.  In 1994, the student programmers left NCSA to form Mosaic Communications Corporation, which became Netscape Communications Corporation.

### B. Cookies

Netscape developed cookies for their first web browser.  At the time, Netscape was under extreme pressure to rapidly develop a browser.  They knew their existence depended upon being first.[11]  To gain adoption, Netscape decided to emphasize security, commerce, and performance in its web browsers and servers.[12]  One of the most important innovations Netscape developed was cookies.  Cookies allowed web sites to maintain information about their visitors.  However, with this ability came a host of privacy and security implications.

---

[8] *Id.* at 238.
[9] *Id.*; *See* also Alan Deutschman, *Imposter Boy*, GENTLEMAN'S Q. (Jan. 1997) (arguing that the idea for NCSA Mosaic belongs to Dave Thompson) *available at* http://www.chrispy.net/marca/gqarticle.html.
[10] GILLIES & CAILLIAU, *supra* note 5, at 241.
[11] Interview by David K. Allison with Marc Andreessen, Founder and Former Chief Operating Officer, Netscape Communications in Mountain View, Cal. (Jun. 1995) *available at* http://americanhistory.si.edu/csr/comphist/ma1.html.
[12] JIM CLARK, NETSCAPE TIME: THE MAKING OF A BILLION –DOLLAR START UP THAT TOOK ON MICROSOFT 109, 157 (1999).

In early web browsers, the Internet was a stateless place.[13] A stateless web is analogous to a vending machine. It has little regard for who you are, what product you are asking for, or how many purchases you have made. It has no memory. This makes commerce difficult. For example, without a state mechanism you would have to buy goods at Amazon.com as you would from a vending machine. You could not buy more than one product at a time and there would be no automatic one-click shopping feature. To overcome this problem Netscape developed Persistent Client State HTTP Cookies.[14] Cookies contain information that allow web sites to maintain information on its users. The first use of cookies was by Netscape to find out if visitors to the Netscape web site were repeat visitors or first time users.[15]

During this time, the Internet Engineering Task Force (IETF) decided to create a standard for state management on the web. Originally, the work was based upon a different technology than cookies that was more sensitive to privacy.[16] Over time, the IETF switched to the Netscape specification. This was largely because the Netscape standard was a working model that was ubiquitous and a de facto standard. However, the standards process soon ran into problems. The IETF felt that Netscape's implementation of cookies was fraught with privacy and security problems.[17]

---

[13] The web was not originally designed to keep state information on the client, *see* Tim Berners-Lee, *HyperText Transfer Protocol Design Issues*, *at* http://www.w3.org/Protocols/DesignIssues.html (last visited Sep. 24, 2001).
[14] Interview with Lou Montulli, Co-Inventor of Cookies, in Bloomington, Ill. (Aug. 2, 1999); Netscape, *Persistent Client State HTTP Cookies*, *at* http://home.netscape.com/newsref/std/cookie_spec.html (last visited Sep. 19, 2001) (the original Netscape specification on cookies); SIMON ST. LAURENT, COOKIES (1998) (providing a good overview of cookies and how to use them).
[15] Netscape's web browser's default home page is Netscape's web site. Thus every user would visit Netscape's web site at least once, *see* Alex S. Vieux, *The Once and Future Kings,* RED HERRING, (Nov. 1 1995) *available at* http://www.redherring.com/index.asp?layout=story&channel=70000007&doc_id=1590015959.
[16] The original basis of the IETF's effort was Kristol's State-Info proposal. Kristol's proposal did not allow for the persistence of cookies and instead the state information would only last a browser session. In contrast, for Netscape's cookies there is no requirement that cookies be destroyed at the end of the browser session, *see* David Kristol, *Proposed HTTP State-Info Mechanism, at* http://portal.research.bell-labs.com/~dmk/session.txt (Aug. 25, 1995).
[17] John Schwartz, *Giving the Web a Memory Cost Its Users Privacy*, N.Y. TIMES, Sep. 04, 2001 *available at* http://www.nytimes.com/2001/09/04/technology/04Cook.html.

The most serious problem with cookies is third party cookies. The intent of Netscape's cookies specification was to only allow cookies to be written and read by the site the person was visiting. For example, if the New York Times placed a cookie on a computer, Amazon.com could not read or modify the New York Times cookie. The purpose of this was to provide security and privacy by only allowing access to information that the site authored. However, Netscape's cookies specification allowed components of a web page to place cookies. A new breed of businesses, the online advertising management companies, has exploited this loophole.

Many sites contract out their banner advertising to advertising management companies. These companies find advertisers for web sites and ensure that the banners appear on the web site. For example, DoubleClick sells advertising space on sites such as ESPN and the New York Times. DoubleClick is also responsible for placing the banner advertising on their clients web sites. Through the loophole of third party cookies, DoubleClick uses its advertising banners on an ESPN web page to place a cookie when a person visits ESPN. DoubleClick can then read and write to that same cookie when the same person visits the New York Times web site. DoubleClick can now aggregate the information about a person's web surfing from its client web sites. They can then create a detailed profile of a person's surfing habits. This has obvious privacy implications.

The cookies standard by the IETF has been critical of third party cookies and requires web browsers to refuse third party cookies by default.[18] Thus far the advertisers who depend on third party cookies, such as Doubleclick and Engage, have been able to ensure the acceptance of third party cookies. The latest version of web browsers by Netscape and Microsoft still accept third party cookies by default. Moreover, the advertisers have promoted new policies such as using privacy policies to continue to allow their third party cookies, despite the obvious security and privacy implications.[19]

### C. Apache

NCSA developed both a browser for viewing pages and server software for delivering web pages. The web server, HTTPd, was based upon the CERN server code. NCSA released the program and its source code for free. Consequently, HTTPd quickly became the most popular web server for the Internet. Many sites chose the free NCSA HTTPd server over Netscape's web servers that cost several thousand dollars.[20]

When HTTPd was first released, the programmers at NCSA quickly patched any problems or bugs that they received. But by 1995, the original team of programmers had left NCSA, and HTTPd was not being updated in a timely manner.[21] This led individuals to begin to "patch" problems they found. An example of a patch was the addition of password authentication by Brian Behlendorf. Other patches improved the security and performance of HTTPd.[22]

Eventually, there were a number of patches for HTTPd circulating across the Internet. However, if someone wanted the benefit of these patches, they would have to download the latest version of HTTPd, and then manually apply all the latest patches.[23] This prompted users of HTTPd to consider updating NCSA's code. According to Østerlie, the individuals saw themselves as disgruntled customers. They were simply filling the gap left by the departure of NCSA's original programmers to Netscape.[24]

---

[18] The standard requires states that third party cookies must not be allowed. It does allow an exception if the program wants to give the user options, however, the default must be set to off, *see* D. Kristol & L. Montulli, *HTTP State Management Mechanism*, RFC 2965, *available at* ftp://ftp.isi.edu/in-notes/rfc2965.txt (Oct. 2000).
[19] The advertising management companies have long proposed methods to continue allowing third party cookies from "trusted cookies" to the use of privacy policies and P3P technology, *see* Daniel Jaye, *HTTP Trust Mechanism for State Management, available at* http://www.ietf.org/proceedings/98aug/I-D/draft-ietf-http-trust-state-mgt-02.txt (Sep. 4, 1998) (trusted cookies); Stefanie Olsen, *IE 6 beta pushes ad networks on privacy, available at* http://news.cnet.com/news/0-1005-200-6285910.html (Jun. 15, 2001) (stating that Microsoft Internet Explorer 6 will require ad networks to use P3P compatible privacy policies in order to place third party cookies).
[20] GLYN MOODY, REBEL CODE 125 (2001).
[21] Andrew Leonard, *Apache's Free-Software Warriors*, Salon, Nov. 20, 1997, *at* http://www.salon.com/21st/feature/1997/11/cov_20feature.html.
[22] *Id*.

Behlendorf began to contact other programmers.[25] By February 1995 the group put together a mailing list called new-httpd and began circulating patches.[26] The project was named Apache, after all, the joke name for the server was "A PatCHy server". The goal of the project was to fix the existing problems and to add enhancements to the server. A key feature of this project was the commitment to keeping the server code available to anyone for free.[27] This commitment to keeping the code free, places Apache as an early exemplar of the open source movement. Today Apache is widely considered a high quality server and is the most popular web server used on the Internet.[28] Its success is used to uphold the open source method of development.

*D. PICS*

The history of PICS begins with proposed legislation to regulate indecent speech on the Internet by Senator Exon in the summer of 1994.[29] Senator Exon reintroduced his legislation in February 1995, and it would eventually become the Communications Decency Act (CDA).[30]

---

[23] Thomas Østerlie, Evolutionary Systems of Innovation, *available at* http://www.pvv.ntnu.no/~toaster/writings/thesis/book/book1.html (last visited Sep. 19, 2001).
[24] *Id*.
[25] Leonard, *supra* note 21. For further background on the core contributors, *see* Patricia Krueger & Anne Speedie, Web Crawlers, Wide Open News, *at* http://www.wideopen.com/story/285.html (Dec. 16, 1999).
[26] MOODY, *supra* note 20, at 127.
[27] Leonard, *supra* note 21.
[28] Aaron Weiss, *Open Source Moves To The Mainstream*, Apr. 10, 2000, *available at* http://www.informationweek.com/781/open.htm (noting the open source Apache's share of the market).
[29] 140 CONG. REC. S 9745 (1994) (including the amendment to S. 1882 by Senator Exon) *available at* http://www.eff.org/Censorship/Internet_censorship_bills/exon_s1822.amend.
[30] Communications Decency Act of 1995, S. 314, 104th Cong. (1995).

This proposed legislation was followed by the now infamous Time cover story on cyberporn.[31] This combination of media and political pressure threw the upstart Internet companies into action.

Microsoft, Netscape, and Progressive Networks created the Information Highway Parental Empowerment Group (IHPEG) in July 1995 to develop standards for labeling content.[32] IHPEG was chosen over the World Wide Web Consortium (W3C) because the members of IHPEG didn't believe the W3C could act quickly enough.[33] At the time, the W3C was a newly emerging consortium seeking to develop web technologies.

In August of 1995 the W3C began planning the development of the Platform for Internet Content Selection (PICS). According to Berners-Lee, the Director of the W3C, "the PICS technology was created specifically in order reduce the risk of government censorship in civilized countries. It was the result of members of the industrial community being concerned about the behaviour of government."[34] The threat of government censorship was tangible because of worries about cyberporn and the CDA. Soon after, the W3C was able to persuade IHPEG to join in the PICS efforts.[35]

By November of 1995 the PICS technical subcommittee released for public review the PICS specifications. PICS could limit access through two methods. First, web sites could self rate their content. They could attach labels that indicated if content contained nudity or violence.

---

[31] Philip Elmer-Dewitt, *On a Screen Near You: Cyberporn*, TIME, Jul. 3, 1995, *available at* http://www.time.com/time/magazine/archive/1995/950703/950703.cover.html.
[32] Information Highway Parental Empowerment Group, *Leading Internet Software Companies Announce Plan to Enable Parents to "Lock out" Access to Materials Inappropriate to Children*, *available at* http://www.csse.monash.edu.au/~lloyd/tilde/InterNet/Law/1995.parental.html (Jun. 15,1995) (noting the members of the IHPEG).
[33] Interview with James Miller, Designer for PICS, in Bloomington, Ill. (Aug. 13, 1999).
[34] Tim Berners-Lee, *Philosophy of the Web - Filtering and Censorship*, *at* http://www.w3.org/DesignIssues/Filtering (last visited Sep. 19, 2001) (detailing Berners-Lee views on filtering and specifically PICS).
[35] W3C, *Industry and Academia Join Forces to Develop Platform for Internet Content Selection (PICS)*, *available at* http://www.w3.org/PICS/950911_Announce/pics-pr.html (Sep. 11, 1995).

Secondly, the PICS specification supports the establishment of third party labeling bureaus to filter content. For example, the Simon Wiesenthal Center could operate a service that filtered out neo-Nazi hate sites. This allows the filtering of web sites without relying on self-rating. In December 1996, the W3C made PICS an official "recommendation", the highest recognition a standard can receive by the W3C.[36]

The final version of the CDA was signed into law on February 8, 1996. Immediately, a lawsuit was filed seeking to overturn the CDA.[37] Albert Vezza, Chairman of the W3C, testified at the trial. His testimony concerned the use of PICS as a method for content selection.[38] The judges were very interested in Vezza's testimony, especially his conclusions that the web has developed almost entirely because the government has stayed out of the way.[39] PICS was held up as an example of how the industry was developing solutions for the problem of access to indecent content by children. The plaintiffs presented PICS technology as a less restrictive alternative to the outright banning of indecent speech on the Internet. The testimony on PICS was persuasive and on June 26, 1997 the Supreme Court found the CDA unconstitutional. The Court noted that the CDA's burden on adult speech "is unacceptable if less restrictive alternatives would be at least as effective in achieving the Act's legitimate purposes."[40]

---

[36] W3C, *W3C Issues PICS as a Recommendation*, *available at* http://www.w3.org/Press/PICS-REC-PR.html (Dec. 3, 1996).
[37] Janet Kornblum & Rose Aguilar, *Clinton signs telecom bill*, CNET NEWS.COM, Feb. 8, 1996, *available at* http://news.cnet.com/news/0,10000,0-1005-200-310586,00.html. For the text of the CDA *see* http://www.epic.org/free_speech/CDA/cda.html.
[38] Testimony of Dr. Albert Vezza at the CDA Trial in Philadelphia (1996) *available at* http://www.ciec.org/transcripts/April_12_Vezza.html (last visited Sep. 20 2001).
[39] Citizens Internet Empowerment Coalition, *Trial Update No. 9*, *at* http://www.ciec.org/bulletins/bulletin_9.html (Apr. 13, 1996).
[40] Reno v. American Civil Liberties Union, 117 S.Ct. 2329, 138 L.Ed.2d 874 (1997).

III. Influences in the Development of Code

There are a variety of influences on the development of code. This part considers the role of political, social, economic, and legal factors. We focus on the entire process of design, development, and implementation of code. While these can be analytically different, our research found these phases intertwined.[41] This section begins with the analysis of universities. This is continued to firms, the open source movement, and consortia. Table 1 summarizes the results of this part and the next section on the review process for code.

Table 1. Summary of Influences and Review Process on Institutions

| *Institution* | *Influences* | *Review Process* |
|---|---|---|
| *University* | Autonomy of researchers<br>Limited resources<br>Desire for peer recognition | Design decision and process is private<br>Public review of code |
| *Firm* | Consumer demand<br>Affected by pressure from politicians and the media<br>Can be affected by government | Decision and process is private and<br>Criterion is based upon economics |
| *Open Source* | Volunteer developers<br>Lack of economic and political pressure<br>Problem of the lowest hanging fruit | Design process is public<br>Design decision is restricted to contributors |
| *Consortium* | Member firms<br>Affected by pressure from politicians and the media<br>Structure of the consortium | Members are usually decision makers<br>Depends upon its structure |

*A. Universities*

The NCSA Mosaic case study found a number of influences that affect the development of code in a university. These influences affected not only NCSA but also CERN. We treat

---

[41] For example, consider the changes to NCSA Mosaic between the first beta release by Andreessen to a final 1.0 release. The implementation process involved considerable feedback from users, which in turned changed the original design of the software by fixing bugs and adding enhancements, *see generally* Ian Sommerville, Software Engineering (5th edition) 210-212 (1995) (the design of software is an iterative process based upon feedback from earlier designs).

CERN as a university style of institution. CERN's structure and motives as a government sponsored basic research laboratory is akin to a university. The first notable influence on the development of code was the autonomous research environment. This provides considerable discretion to researchers for the development of code. The next influence focuses on the limited resources at universities. The lack of resources affects both the content of code as well as the process of development. The third influence is the desire for peer recognition by the developers.

The first influence is the autonomy and freedom given to developers within university research environment. The result is pioneering code. There is a direct link between the freedom of researchers and the innovative code produced in a university setting. One study found that the innovative research of Nobel Prize winners depends upon an institutional environment that is characterized by freedom.

> not absolute freedom, and not endless time and boundless resources, but freedom above all to use one's own personality in pursuit of a scientific objective, freedom to pursue hunches down possibly pointless avenues of exploration and freedom to theorize, experiment, accept, or reject, according to the principal investigator's own judgment, with no interference.[42]

The case studies show that NCSA and CERN allowed their researchers considerable freedom in their projects. CERN was an institution devoted to fundamental research with little regard to the potential profits of innovative results. CERN's computing environment promoted the development of new software from running the coke machine to software for a physics experiment.[43] Within this institutional environment, Berners-Lee was allowed to work on his radical proposal for creating a networked hypertext program, the World Wide Web. Similarly, NCSA Mosaic was developed in an academic environment that gave Andreessen considerable autonomy.

---

[42] JOHN HURLEY, ORGANISATION AND SCIENTIFIC DISCOVERY 4 (1997).
[43] BERNERS-LEE, *supra* note 6, at 43.

The next factor is the limited resources within universities. The lack of resources affects the code's content as well as the process of developing code. The lack of resources can lead university researchers to focus on developing the building blocks and standards for future work. So instead of developing a large complex program, a university researcher may instead focus on demonstrating that such a program could be completed. This was illustrated during the development of the web. Berners-Lee lacked the resources to develop web browsers for the major computing platforms. So instead he focused on developing standards and reusable building blocks of code. These blocks of code were known as libwww and became the basis of future web browsers and servers.[44]

The lack of resources affects the process of developing code. For example, projects at universities can't depend upon a large staff. The consequence is that functions such as technical support and documentation are viewed as extras and therefore not fully supported. Moreover, the lack of resources gives researchers the impetus to seek resources outside the university. This was evident in the development of the web. Berners-Lee did not have a staff inside CERN and began to look for development help outside CERN.[45] He was able to persuade university students to develop web browsers. Erwise, the first browser for X-Windows, was written by students from Helsinki University in 1992.[46] Thus the limited resources affected the process of the development of the web.

The third notable influence is peer recognition. This motivation governs academic research. Academics aspire to have their work cited by others or have their new tool or technique adopted by their peers. A consequence of peer recognition is that code is produced with little regard to economic gains. Instead, code is designed within universities to generate interest among their peers.

---

[44] GILLIES & CAILLIAU, *supra* note 5, at 209.
[45] BERNERS-LEE, *supra* note 6, at 46.
[46] *Id.* at 55-56.

The desire for peer recognition among NCSA's student programmers was manifested in making "cool" programs. For example, Andreessen decided to design NCSA Mosaic browser to view images. He thought this would be cool, because it made the web more attractive and easier to use. However, many within the Internet community including Berners-Lee disagreed with Andreessen's decision. Berners-Lee thought of the web as a tool for serious communication between scientific researchers. He didn't think the design of browsers should be about what looks cool.[47] This example shows the influence of peer recognition and also its lack of uniformity. In this case, Berners-Lee and Andreessen sought peer recognition from two different groups.

The desire for peer recognition extends to the institutional level. This is unusual. Often influences at an individual level are not the same at the institutional level. The consequence is that universities promote and support code that can enhance their reputation. This occurred during the development of NCSA Mosaic. Once NCSA understood its significance, as indicated by downloads, NCSA devoted more resources to the development of Mosaic.[48] The University of Illinois also began touting the accomplishments of NCSA Mosaic. It used the prestige of NCSA Mosaic to enhance its own status.[49] Thus the desire for peer recognition affects both researchers and their institutions.

---

[47] JOHN NAUGHTON, A BRIEF HISTORY OF THE FUTURE: FROM RADIO DAYS TO INTERNET YEARS IN A LIFETIME 244-245 (2000).
[48] Department of Computer Science University of Illinois, *Illinois's Boys Make Noise: And they're doing it with Mosaic*, COMPUTER SCIENCE ALUMNI NEWS, *available at* http://www.cs.uiuc.edu/whatsnew/newsletter/winter94/mozilla.html (Winter 1994).
[49] *See* University of Illinois, Facts 2001, *available at* http://www.admin.uiuc.edu/pubaff/facts96.html (last visited Sep. 2, 2001) (highlighting the role of Mosaic in a background statement about the university)

*B. Firms*

Firms produce goods and services which they expect people to buy.[50] An important consideration for firms is the anticipation of consumer needs. These economic concerns are the primary motivator of firms and the efficacy of this process surrounds us everyday. This section largely focuses on how firms anticipate consumer demand and the consequences that flow from this. The final point is how firms can be affected by pressure from the media and politicians.

The major influence on firm is the anticipation of consumer demand. Firms need to correctly anticipate and create consumer demand to generate revenue. If they cannot meet consumer demand they will cease to exist. For example, Netscape anticipated demand for new browsers and servers to support commerce. To support commerce, Netscape added new proprietary standards such as cookies and the Secure Sockets Layer.[51] These technologies were widely supported and consequently Netscape to gained a large share of the market.[52]

The focus on consumer demand can lead to firms missing innovative changes in technology. Firms will not invest in uncertain or unproven technology without a commensurate rate of return. This leads to underinvestment in basic research or radical new inventions.[53] This is illustrated in the development of the web. After Berners-Lee conceived of the web, he approached a number of firms that had built hypertext products. He encouraged them to extend their software to the Internet through the web concept. But none of them were interested in his vision. They didn't think there was any money to be made.[54]

---

[50] Our definition of firm goes beyond the strict legal definition of corporation and is meant to encompass other forms of individual cooperation in the marketplace such as joint ventures, associations, or cooperatives. We also treat corporate research laboratories as firms, because of the recent trend that emphasizes applied research over basic research in these laboratories, *see* NATIONAL ACADEMY OF SCIENCE, MAKING IT BETTER: EXPANDING INFORMATION TECHNOLOGY RESEARCH TO MEET SOCIETY'S NEEDS 72-73 (2000); (noting how IBM, AT&T and Lucent Laboratories, and Xerox laboratories have redirected their research to meet business interests).
[51] ROBERT H. REID, ARCHITECTS OF THE WEB: 1,000 DAYS THAT BUILT THE FUTURE OF BUSINESS 35 (1997). Secure Sockets Layer allows for encrypted communications between a browser and server.
[52] David Legard, *Microsoft Wins Browser Battle,* PC WORLD, *available at* http://www.pcworld.com/news/article.asp?aid=13697 (Nov. 09, 1999) (noting Netscape's had over 70% of the browser market in the late 1990s).
[53] Richard R. Nelson, *The Simple Economics of Basic Scientific Research*, 67 J. POL. ECON. 297 (1959).

The focus on consumer demands ignores the needs of citizens or society. Firms are not concerned about social values that are deemed unprofitable. Despite the fact, that these values may be important to citizens. In the case of cookies, Netscape was not going to spend its resources developing unprofitable code to minimize the privacy loss of cookies. This explains why early versions of Netscape contained no cookie management tools or even documentation about cookies. Security is another value that is deemed unprofitable. Firms are ignoring security, because firms do not think that improved security will increase their market share. Thus the lack of security is a deliberate business decision in some products.[55] As a result, there has been a rash of security problems for the Internet that affects everyone.[56] This neglect of society's long-term interests by firms is natural. This calls for government intervention to provide incentives for firms to consider general social values.

Finally, the case study of cookies shows how firms are affected by political and media pressure. Firms react to political and media pressure. This is another influence on the behavior of firms. This was evident in our cookies case study when the media uproar over online privacy problems led to hearings by the Federal Trade Commission (FTC). The hearings only touched the surface of the privacy issues and technical features. But at both hearings, Netscape was forced to discuss how cookies work and how Netscape would improve privacy. As a result, the government hearings pushed the browser makers to provide some cookies management tools and to improve documentation of cookies.[57]

---

[54] BERNERS-LEE, *supra* note 6, at 26-28. This is not unique to this case study. During the development of the Internet, AT&T ridiculed the concept of "packet based" communication, which the Internet would later be based upon. AT&T didn't see any reason for such a new communication method and actually refused to allow "their" network to carry such communication even though the U.S. Government would have funded the research, *see* KATIE HAFNER & MATTHEW LYON, WHERE THE WIZARDS STAY UP LATE: THE ORIGINS OF THE INTERNET 64 (1996).
[55] Robert X. Cringely, *The Death of TCP/IP: Why the Age of Internet Innocence is Over*, Aug. 2, 2001 *available at* http://www.pbs.org/cringely/pulpit/pulpit20010802.html.

*C. Open Source*

The open source movement is primarily influenced by its volunteer developers. The first section notes the role of volunteer developers for the open source movement. The second section argues that a consequence of volunteer developers is that the media, politicians, and economics provide little influence on the development of code. At times, the open source movement can even counter dominant political or economic concerns. Finally, another consequence of volunteer developers is the problem of the lowest hanging fruit. This leads to a lack of work on projects deemed uninteresting.

The primary influence on open source code is its volunteer developers. It is the developers who decide what code will be written and on what schedule.[58] This can be useful since it collapses the distinction between users and developers. The developers don't have to envision an imaginary user.[59] This reduces a problematic step in the development process.[60]

---

[56] Scott Charney, *The Internet, Law Enforcement and Security*, Internet Policy Institute, *available at* http://www.internetpolicy.org/briefing/charney.html (last visited Sep. 20, 2001) (arguing government has a role for online security and therefore not relying solely on the private sector).

[57] In June of 1996 the FTC held a Public Workshop on Consumer Privacy On the Global Information Infrastructure. At this hearing Harter of Netscape announced that the next version of Netscape (version 3.0) would allow users an option to be alerted whenever a cookie is placed on their computer. At the 1997 FTC Workshop, Harter noted in version 4.0 the user sees the following cookie choices: Accept all cookies; Accept only cookies that get sent back to the originating server; Disable all cookies; Warn me before accepting a cookie, *see* Netscape, *FTC: Consumer Privacy Comments Concerning the Netscape Communications Corporation*, P954807, Jun. 6, 1997, *available at* http://www.ftc.gov/bcp/privacy/wkshp97/comments2/netscape.htm. Recently, Microsoft touted its improvements to cookie management at a Senate hearing into privacy, *see* Senate Committee on Commerce, Science & Transportation, Need for Internet Privacy Legislation, July 11, 2001 *available at* http://www.senate.gov/~commerce/hearings/hearings.htm (last visited Sep. 20, 2001)

[58] This is well stated by Jordan Hubbard, a founder of the open source FreeBSD project:
> Developers are very expensive commodities (just ask any IT hiring manager) and getting their expensive time and effort for free means that it comes with certain stipulations. The developer has to have a personal interest in the features in question and they will implement those features according to the features in question, and they will implement those features according to the demands of their own schedule, not anyone else's. An interview with Jordan Hubbard, *available at* http://www.workingmac.com/article/32.wm (Aug. 16, 2001).

The development is biased by the motivations of these developers. Most open source software is motivated by utilitarian concerns. That is "every good work of software starts by scratching a developer's personal itch."[61] In the case of Apache, the developers were trying to improve a web server that they used. This was a purely utilitarian concern.

The influence of economic and political pressure on the open source movement is minimal, because it is led by an international team of volunteers. These volunteer developers are focused on what is interesting for them. This allows the development of code with features that contain little economic value. For example, Mozilla, an open source browser based on Netscape, contains features such as cookies management, and the ability to block images from third party web sites as well as pop up advertising windows.[62] These features are there because the open source community feels they are important.

At times, the open source movement can be defiant to economic and political influences. It was the open source movement that refined and distributed the decentralized file sharing system Gnutella. Moreover, the open source movement is now attempting to create an anonymous decentralized file sharing system. This system, Freenet, will make it impossible for governments to track down users or remove information.[63]

---

[59] There are design approaches that involve the user, such as participatory design, *see* DOUGLAS SCHULER & AKI NAMIOKA, PARTICIPATORY DESIGN: PRINCIPLES AND PRACTICES (1993) (leading textbook on participatory design); TERRY WINOGRAD, BRINGING DESIGN TO SOFTWARE (1996) (how to use participatory design to improve the development of software).
[60] Paul Quintas, *Software by Design*, i*n* COMMUNICATION BY DESIGN: THE POLITICS OF INFORMATION AND COMMUNICATION TECHNOLOGIES 93 (Robin Mansell & Roger Silverstone eds. 1996).
[61] Eric S. Raymond, *The Cathedral & the Bazaar: Musings on Linux and Open Source by an Accidental Revolutionary*, *available at* http://www.tuxedo.org/~esr/writings/cathedral-bazaar/cathedral-bazaar/.
[62] *See* Sneak Peak: Netscape 6 Preview Release 1, CNET NEWS.COM *available at* http://www.cnet.com/internet/0-3779-7-1581725.html (improved cookies management features); *Banners, be gone*!, MOZILLA WEEK, *at* http://www.netartmagazine.com/mozweek/archives/00000026.html (Mar. 9, 2001) (how to block images from third parties with Mozilla); *Mozilla 0.9.2 Released*, BROWSERWATCH *at* http://browserwatch.internet.com/news/stories2001/news-20010629-1.html (Jun. 29, 2001) (blocking pop up windows).

A third influence is the problem of the lowest hanging fruit. This problem acknowledges that volunteers want to work on interesting tasks. Unlike in firms, there is no leader who can command or require anyone to do anything. This problem is endemic in open source projects and is described accordingly:

> Those who can program naturally tend to work on programs they find personally interesting or programs that looks cool (editors, themes in Gnome), as opposed to applications considered dull. Without other incentives other than the joy of hacking and "vanity fair" a lot of worthwhile projects die because the initial author lost interest and nobody pick up the tag.[64]

Thus this leads to projects which developers think are interesting, such as a C complier or an mp3 player. The consequences are that developers may not work on code that is in greater demand or more socially beneficial.

This problem occurred during the development of Apache. At one point, work on Apache dramatically slowed. Østerlie notes this crisis occurred when the work before the group were of a menial kind. According to Østerlie, "because of the difficulty of integrating new features with the existing NCSA code base, the "lowest hanging fruits" had been picked. Thus in order to get further with enhancing the web server a total rewrite would be necessary. This would be a difficult and mundane task."[65]

The problem of the lowest hanging fruit was also manifested during the development of web browsers. The development of web browsers, such as NCSA Mosaic and Erwise, relied on volunteer programmers all across the world. According to Berners-Lee, these developers were more interested in "putting fancy display features into the browsers—multimedia, different colors and fonts—which took much less work and created much more buzz among users.[66] Berners-Lee wanted the developers to focus on a much more substantive issues—the addition of editing features to the browser. The concept of a browser/editor was important for Berners-Lee. He envisioned the web as a place where it should be as easy for people to publish as it is to read. Berners-Lee believes that the reason people focused on browsing over writing and editing features was that it just wasn't fun to create an editor.[67] Thus the problem of the lowest hanging fruit can bias the development of code towards what is interesting for developers.

---

[63] John Markoff, *The Concept of Copyright Fights for Internet Survival*, N.Y. TIMES, May 10, 2000, *available at* http://www.nytimes.com/library/tech/00/05/biztech/articles/10digital.html (discussing the Freenet project).
[64] Nikolai Bezroukov, *Open Source Software Development as a Special Type of Academic Research (Critique of Vulgar Raymondism)*, FIRST MONDAY, Oct. 1999 *at* http://www.firstmonday.dk/issues/issue4_10/bezroukov/.
[65] The problem of the lowest hanging fruit should not be confused with the most easily picked fruit. The distinction is not the ease of the project but rather whether the projects were one that were intriguing and interesting to developers, *see* Østerlie, *supra* note 23.

### D. Consortia

A consortium consists of a number of firms engaged in cooperative research and development. Their rationale is to fund research that is useful to all of them and would not otherwise be developed by a single firm.[68] The work might not be completed by one firm because of the sheer cost or the need for a standard which competing firms will adopt. As a result, the largest influence on the development of code within a consortium is its member firms. In our case study, the primary motivation for PICS was the result of pressure from the media, politicians, and the law on the members of the W3C. Secondly, the structure of the W3C influenced the development of PICS.

---

[66] BERNERS-LEE, *supra* note 6, at 71.
[67] *Id.* at 57.
[68] If there are strong economic incentives for certain code, this work will be done outside the cooperative reaches of a consortium. An example of a successful consortium is Sematech which is devoted to supporting the semiconductor industry in the United States, *see* LARRY D. BROWNING & JUDY C. SHETLER, SEMATECH: SAVING THE U.S. SEMICONDUCTOR INDUSTRY (2000).

A consortium allows a number of firms to work together to develop a technical solution useful to all of them. One use of a consortium is to develop a technology to prevent government regulation and allay the fears of politicians. This was illustrated in the PICS case study. PICS developed in response to pressure from the media, politicians, and the CDA on the members of the W3C. According to James Miller, cochairmen of the PICS Technical Committee, PICS was motivated by desires to avoid regulation. Miller remarked that, "if we hadn't had the bill going through congress [the CDA] there is no way this group would have come together, in fact its evidenced by the fact we had been asked at our previous members meeting by both IBM and AT&T to look into this, and nothing had happened." The pressure led to the W3C placing PICS on a rapid development cycle. As Miller revealed, the timetable for the development of PICS was set by lawyers challenging the constitutionality of the CDA. And when the CDA was found unconstitutional, the development of software for PICS was essentially abandoned.[69]

The structure of a consortium also influences the development of code. A closed private process is consistent with consortia that strive for the rapid development of standards. This was used by the W3C in the development of PICS. The team consisted of a dozen people largely from the consortium's membership. The team communicated privately during the entire developmental process.[70] This allowed them to complete their work in a matter of weeks.

A consequence of the consortia's structure is the inadequate consideration of the needs of entities who are not members, such as independent software vendors and end users. This can result in ineffective and technically poor solutions.[71] This occurred in the development of PICS. PICS's system of third party labeling bureaus was not based upon a sustainable economic model for filtering software vendors. Consequently, PICS is of little use to firms selling filtering software to libraries and parents. Similarly, the assumption by the PICS team that most people would self-rate the content of their pages was totally unreasonable. These are understandable consequences, because the needs of end users and the commercial filtering firms were not addressed.

---

[69] Today, PICS rests upon web users and web sites labeling their own pages. There is no server software to operate third party labeling bureaus for PICS, *see* infra note 85.
[70] Their communications have not been made public.
[71] S*ee* Andrew Updegrove, *Standard Setting and Consortium Structures*, STANDARDVIEW, Dec. 1995, 145.

The consortium approach formulated in PICS is being attempted again in the context of privacy with Platform for Privacy Preferences (P3P) project. Once again, a consortium appears as a natural solution to political pressure. The consortium approach is trying to head off government regulation by claiming an industry solution or self-regulation. In the case of P3P, the W3C has worked with industry to ensure that P3P will be widely adopted by the software vendors, such as Microsoft. However, in doing so, they have neglected the end user and built a product that reflects the industry's view of privacy and not the expectations of most people.[72]

## IV. THE REVIEW PROCESS

This section discusses the review process. There are several important elements in this decision. Who gets to make the decision? What are the criteria? How public is the decision-making process? These decisions are important, because they decide what features are included in the code as well as when code is disseminated. We begin by discussing the review process at the university and then the firm. The third section focuses on the role of the Internet Engineering

---

[72] Christopher D. Hunter, *Recoding the Architecture of Cyberspace Privacy: Why Self-Regulation and Technology Are Not Enough*, International Communication Association Convention, Acapulco, Mexico, *available at* http://www.asc.upenn.edu/usr/chunter/net_privacy_architecture.html (Jun. 2000). For background on P3P as well as the relationship between the technical and policy decisions, *see* Lorrie Faith Cranor and Joseph Reagle, *Designing a Social Protocol: Lessons Learned from the Platform for Privacy Preference, in* TELEPHONY, THE INTERNET, AND THE MEDIA (Jeffrey K. MacKie-Mason & David Waterman eds., 1998) *available at* http://www.research.att.com/~lorrie/pubs/dsp/. *See also* the W3C page with background info on P3P *at* http://www.w3c.org/P3P/.

Task Force. This institution is important for the Internet and has a different review process. We conclude by examining the review process for the open source movement and consortia.

*A. University*

In a university research project, it is the developer or the head of a project that decides whether to release code and the substance of the code. The decision-makers and their criteria for the review process are private. This is the result of the autonomy and freedom within the university. The lack of external pressure allows universities to design and develop code that is risky and may fail. This risk taking is important for pushing the boundaries of knowledge and creating innovative products. In our case study, it was Andreessen and Berners-Lee who decided what features to include in their browsers. They were the ones who announced the availability of the latest versions of their browsers on the Internet.[73] Thus NCSA and CERN provided their researchers considerable autonomy in the review process for code.

The decision to disseminate code publicly is natural within a university. This is consistent with the academic norms of openness and the need to publish research. These norms support the involvement of public comment and criticism once the initial work is completed.[74] This is how Berners-Lee and Andreessen acted. They both released their code publicly through the Internet and asked for feedback. They considered the public as their customers.[75] Thus the university allows its researchers autonomous decision-making power during the review process but expects them to disseminate their code publicly.

---

[73] Marc Andreessen, *NCSA Mosaic for X 0.9 released*, WWW-TALK MAILING LIST, *available at* http://www.webhistory.org/www.lists/www-talk.1993q1/0227.html (Mar. 4, 1993) (announcing a new beta version of NCSA Mosaic by Andreessen).
[74] Stevan Harnad, *Learned Inquiry and the Net: The Role of Peer Review, Peer Commentary and Copyright*, a*vailable at* http://citd.scar.utoronto.ca/EPub/talks/Harnad_Snider.html (1997) (commenting on the role of peer review).
[75] In February 1993, a message was posted congratulating Andreessen on NCSA Mosaic and asking him why he cared about what others thought, since they weren't customers of NCSA. Andreessen replied:
> Well, you literally are our customer. But that's probably beside the point... we do care what you think simply because having the wonderful distributed beta team that we essentially have due to this group gives us the opportunity to make our product much better than it could be otherwise.

*B. Firms*

Firms are focused on releasing code that is profitable. Profitability is the criterion for releasing code. This leads to a review process in which the decision-makers and the process are private. There is no reason to provide information about potential code to rivals. Moreover, there is tremendous pressure on firms to introduce their code rapidly into the market to gain advantage over competitors. This process was evident in Netscape's behavior. Netscape analyzed the potential market and then decided to develop both a web browser and a web server. They believed the addition of features supporting commerce would be profitable. Furthermore, they understood that their decisions had to be made quickly to gain a competitive advantage.

The criterion of profits was evident in a number of decisions by Netscape. Netscape chose to incorporate cookies into their new browser. They did this despite the potential security and privacy issues. Netscape did not wait for the IETF to define a cookie standard. Instead, they rushed to meet the market expectations. And a few years later, Netscape decided to continue allowing third party cookies. This decision was made with full knowledge of the privacy and security risks, as well as the Internet community's disapproval of third party cookies. Netscape's criterion was its economic interest. This is how firms operate. The consequence is that values that are deemed unprofitable are not factored into a firm's review process.

---

We very much value constructive criticism from knowledgeable people, see Marc Andreessen, *Xmosaic Experience,* WWW-TALK MAILING LIST, a*vailable at* http://www.webhistory.org/www.lists/www-talk.1993q1/0176.html (Feb. 25, 1993).

## C. Internet Engineering Task Force

The next institution of interest is the Internet Engineering Task Force (IETF). We have not explicitly discussed this institution in this paper for two reasons. First, the IETF does not develop code. The IETF develops standards so code can interconnect and interoperate. Secondly, our research found the process and development of standards within the IETF analogous to the open source movement. Anyone may join the IETF, all communication is public, and the final standard is available to the public for free. However, the review process of the IETF is different than that of other institutions and worthy of examination.

Decisions in the IETF are made on the basis of "rough consensus and running code." The decision makers for IETF standards are the public. IETF standards are made on the basis of rough consensus and the public is invited to join. Consider the debate on standards for the Multipurpose Internet Mail Extensions (MIME). The debate included hundreds of people. And when Steve Jobs of Next Computer appealed to Nathaniel Borenstein, the author of the MIME standard, for some changes, Borenstein said no. Borenstein said that he believed it was absurd that "because that you were a famous executive that your opinion should outweigh the Internet community's reasoned debate."[76] This is the ideal of the IETF. The IETF depends upon a general community consensus to determine Internet standards.

The emphasis on running code requires a standard to have at least two working implementations.[77] This is different from official standardization bodies that are not strongly motivated by implementations of standards. Scholars refer to the IETF as a gray standard body, because standards are initiated and driven by implementers.[78]

---

[76] Interview with Nathaniel Bornstein, Author of MIME Standard, in Bloomington, Ill. (Sep. 17, 1999).
[77] Scott Bradner, *The Internet Standards Process -- Revision 3*, RFC 2026, *available at* http://www.ietf.org/rfc/rfc2026.txt (Oct. 1996).

IETF standards are developed publicly. This public development process can allow for the identification of potential defects in the standard. It was the IETF standards process that quickly recognized the privacy and security flaws in cookies such as third party cookies.[79] This illustrates the power of the public review process. The public review process would not let the privacy and security flaws pass unnoticed. Moreover, if cookies were initially placed within the IETF standards process, their privacy and security flaws could have been detected and fixed before the widespread implementation of cookies. As a result we would not have the problems of third party cookies.

*D. Open Source*

The open source movement also follows the Internet adage, "rough consensus and running code." The review process is open to public comment, from which rough consensus is judged. As a result, the open source movement can quickly identify problems in the code. This quality is also found in the IETF review process. For example, during the development of Apache over 3000 people contributed reports on problems with the code.[80]

The number of decision makers can range from democratic to authoritative. In the case of Apache, there was a core group of people who made the decisions. This clique of developers determines the final form of Apache through a voting process.[81] In contrast, other successful open source projects are run in an authoritative manner. For the Linux operating system, it is up to Linus Torvald whether to accept a patch.[82]  While he usually accepts the recommendations of his core team, he does have the discretionary power to do as he pleases.

---

[78] Tineke M. Egyedi, *Instiutional Dilemma in ICT Standardization: Coordinating the Diffusion of Technology, in* INFORMATION TECHNOLOGY STANDARDS AND STANDARDIZATION: A GLOBAL PERSPECTIVE 55 (Kai Jakobs ed., 2000).
[79] Schwarz, *supra* note 17.

The criterion for the review process is not fixed.  In the case of Apache, the criterion focused on the addition useful features and the removal of errors in the code.  In other cases, the lack of economic or political criteria can lead to the inclusion of features that are either politically unpalatable or not in the economic interest of the Internet.  For example, the open source web browser Mozilla is capable of blocking the nefarious pop up windows used for advertisements.[83]

*E. Consortia*

The review process of a consortium depends upon its structure chosen by its membership. The final decision maker for the W3C is its director.  Naturally, he is going to do something that most of the members support, because member support is vital for a consortium.  Thus the decision is not subject to market forces, but the discretion of the consortium's members.  The result is consortia may develop standards such as PICS, which firms have little economic incentive to implement.

Typically, a consortium has a choice between speed and consensus.  The more parties involved in the decision making process, that is consensus, the longer the process of review.  In the case of the W3C, its process is geared towards speed over consensus.  The PICS study shows that within months the W3C can draft a standard and have it in final form in about one year.  In contrast, other standard organizations such as the International Organization for Standardization can take years to develop a standard.[84] The tradeoff is that its members may not support some of these standards. In the case of PICS, only a few firms ever fully supported all the capabilities of PICS in their products.[85]

---

[80] Roy T. Fielding et al., *A Case Study of Open Source Software Development: The Apache Server*, PROCEEDINGS OF ICSE 3-5 (2000) *available at* http://www.bell-labs.com/user/audris/papers/apache.pdf.
[81] Bezroukov, *supra* note 64.
[82] Id.
[83] See *supra* note 62.

V. CONCLUSION

This paper has shown the importance of institutions during the development process of code. We sought to identify different factors that influence the development of code as well as the decision making process to release code. We found these factors vary by institution. As a result, institutions differ in the values that they incorporate into code or communications technology.

The influences that affect code within a university are the autonomy of the researchers, the lack of resources, and the desire for peer recognition. These factors aid in the creation of innovative code such as NCSA Mosaic. Within a firm, the influence on the design of code is the anticipation of consumer demand. A consequence is that firms do not develop code that is deemed unprofitable. But a firm's code can be influenced by pressure from the media and politicians. In contrast, the open source movement's reliance on volunteer developers may lead to code that is unprofitable and politically unpalatable. However, the open source movement's reliance on volunteers can lead to a lack of work on projects that are considered uninteresting. Finally, we noted that the goals and structure of a consortia's are heavily influenced by the demands of its members. It was the desire for the W3C's members to avoid regulation that led to the development of the PICS protocol. Similarly, the structure of the W3C favored a speedy private development process.

---

[84] Roy Rada, *Consensus Versus Speed, in* INFORMATION TECHNOLOGY STANDARDS AND STANDARDIZATION: A GLOBAL PERSPECTIVE 21 (Kai Jakobs ed., 2000).
[85] In tandem with the lack of a business model for public labeling bureaus was the lack of support from software vendors. The server software for creating label bureaus was only developed for a few servers. Most notably, Netscape and Microsoft did not have this feature. The W3C's web page indicates the only commercial server software was IBM, *see* http://www1.raleigh.ibm.com/pics/servers.html. IBM dropped support for PICS when it adopted Apache as its web server, for example *see* http://www-1.ibm.com/servers/eserver/iseries/software/http/services/apache.htm. Although the W3C's Jigsaw server supports PICS, however it is not intended as a production system but instead as a reference system to test standards.

The review process for code differs by institution. These differences included the criteria for dissemination of code, who the decision makers were, and the public nature of the process. Within a university, it is the developers who decide when to release code. They are not subject to any criterion, but are obliged to make their code publicly available. Firms are instead subject to an economic criterion. Consequently, firms will chose their own economic well being over values such as privacy that may be unprofitable. Firms also use private decision-making with an emphasis on speed during the review process. We discussed the role of the IETF's review process, because of the IETF's significance in developing standards for the Internet. The IETF's process allows the public to join in the decision-making process. The IETF places emphasis on running code as its criterion. The consequence of the openness of the IETF and open source movement's review process is that problems can be quickly detected. The decision maker for the open source movement differs from the IETF. The decision makers for the open source movement vary by project from democratic to authoritarian. Finally, a consortium's review process depends upon its structure. The structure can vary from favoring consensus to speed.

The case studies have shown how the influences and review process for Code affect the final features of code.[86] Future work will address how these influences are manifested in the choices of open standards, open source code, choice of intellectual property, marketing, technical support, and bugs in the code. More importantly, this research will consider how to best address general societal problems with institutions. When are problems best solved by government or by a firm? The analysis in this paper will then allow for the development of normative proposals. These proposals will focus on how the government and the private sector can encourage innovative code that is responsive to societal concerns.

---

[86] While these features are the outcome of the institutions, this is not a one-way effect. These features also feedback and affect the development of code. For example, the choice of what sort of intellectual property protection to seek can affects the development of code. Public domain or open source software may lead to outside contributors during the development of code.